# Phase-locked multi-terahertz electric fields exceeding 13 MV/cm at 190 kHz repetition rate

Matthias Knorr[1], Jürgen Raab[1], Maximilian Tauer[1], Philipp Merkl[1], Dominik Peller[1], Emanuel Wittmann[2], Eberhard Riedle[2], Christoph Lange[1] and Rupert Huber[1]

[1]*Department of Physics, University of Regensburg, 93040 Regensburg, Germany*
[2]*LS für BioMolekulare Optik, Ludwig-Maximilians-Universität München, Oettingenstrasse 67, 80538 München, Germany*

We demonstrate a compact source of energetic and phase-locked multi-terahertz pulses at a repetition rate of 190 kHz. Difference frequency mixing of the fundamental output of an Yb:KGW amplifier with the idler of an optical parametric amplifier in GaSe and LiGaS$_2$ crystals yields a passively phase-locked train of waveforms tunable between 12 and 42 THz. The shortest multi-terahertz pulses contain 1.8 oscillation cycles within the intensity FWHM. Pulse energies of up to 0.16 µJ and peak electric fields of 13 MV/cm are achieved. Electro-optic sampling reveals a phase stability better than 0.1 π over multiple hours combined with free CEP tunability. The scalable scheme opens the door to strong-field terahertz optics at unprecedented repetition rates.

Ultrashort pulses in the terahertz (THz) and mid-infrared region of the electromagnetic spectrum have attracted tremendous interest in the past few years as resonant probes of low-energy elementary excitations in condensed matter [1,2]. The combination of CEP-stable pulses with ultrabroadband electro-optic sampling [3-8] has allowed for studies of electronic and structural dynamics of molecules and solids, on time scales faster than a single cycle of the carrier wave [1,2]. The recent advent of high-power sources [9-13] has prompted an ongoing revolution of ultrabroadband THz nonlinear optics and resonant THz quantum control of condensed matter [14-20]. In particular, when the ponderomotive energy exceeds the fundamental bandgap of semiconductors or dielectrics, the carrier wave acts like an AC bias field that can accelerate and recollide quasiparticles [15,16]. It can drive dynamical Bloch oscillations and high-harmonic generation [17], or induce tunneling of electrons out of sharp metal tips [18] or through the tunneling junction of a scanning tunneling microscope (STM) [19,20]. In the multi-THz range, non-perturbative dynamics of this nature, often dubbed 'lightwave electronics', have occurred for field amplitudes typically above 10 MV/cm.

Optical rectification, i.e. difference frequency generation (DFG) within the broad spectrum of a single femtosecond near-infrared (NIR) pulse, gives rise to passively phase-locked THz pulses [3-8,21]. While this concept warrants a particularly stable carrier-envelope phase (CEP), its observed low quantum efficiency has made it a popular choice for the generation of probe pulses [1,2]. Difference frequency mixing between the signal waves of two optical parametric amplifiers driven by the same pump laser, in contrast, has generated CEP-stable few- and single-cycle multi-THz pulses with field amplitudes in excess of 10 MV/cm or even above 100 MV/cm [9-11,22]. An innovative in-line scheme of two-color parametric amplification in a single OPA has further improved the long-term CEP-stability of the transients [13]. Owing to the large required pulse energies, amplifier-based THz sources have been limited to repetition rates in the few-kHz regime. Yet, a broad variety of future applications, such as frequency comb metrology and molecular fingerprinting with solid-based high-harmonic sources, lightwave-STM, or strong-field light-matter interaction with massively improved signal-to-noise ratios, call for substantially increased repetition rates while keeping the field amplitudes above 10 MV/cm. Such sources have been limited to wavelengths shorter than 5 µm [23].

Here, we introduce a table-top light source generating multi-THz few-cycle transients with field strengths exceeding 13 MV/cm at a repetition rate as high as 190 kHz. Our approach exploits DFG of femtosecond NIR pulses from a laser amplifier and the idler output of an OPA pumped by the second harmonic of the laser. The source is tunable between 12 and 42 THz. Its design ensures inherent phase locking while its compact layout using only one OPA provides excellent phase stability.

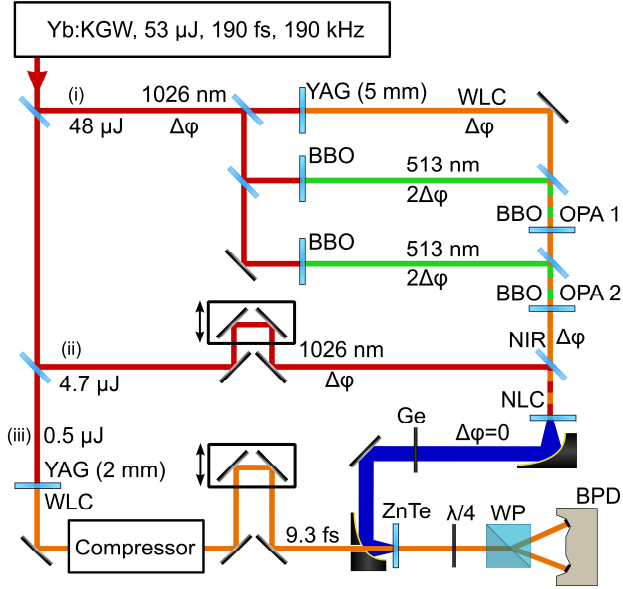

**Fig. 1.** Generation and electro-optic detection of energetic, CEP-stable multi-THz pulses at 190 kHz repetition rate. The pulse energies and the CEP fluctuations are given for each branch, starting with the fluctuation Δφ of the amplifier system. YAG, YAG crystal; BBO, 2-mm BBO crystal; WLC, white light continuum; OPA 1/2, stage 1 and 2 of the SHG pumped OPA; NLC, non-linear crystal for DFG; Ge, 500-μm germanium; ZnTe, 6.5-μm ZnTe crystal; λ/4, quarter-wave plate; WP, Wollaston prism; BPD, balanced pair of photodiodes;

The schematic of the source is sketched in Fig. 1. We start with 190-fs pulses centered at a wavelength of 1026 nm from a commercial regenerative Yb:KGW laser amplifier system (PHAROS 10; Light Conversion) with a repetition rate set to 190 kHz. The pulse energy of 53 μJ is split into three branches serving (i) as the pump of a NIR OPA, (ii) as the pump of a multi-THz DFG stage and (iii) as the generation pulse of a gate for electro-optic detection. The compact OPA encompasses two amplification stages both of which are driven by the second harmonic of the laser. A white-light continuum from a 5 mm YAG crystal [24] seeds the first OPA process in a 1-mm-thick type-II BBO crystal. Subsequently, the idler wave of the first OPA stage is seeded into a second identical BBO crystal to boost the energy of the signal and idler pulses. Typical energies of 4.2 μJ and 3.2 μJ are reached for a signal/idler pair at 900 and 1200 nm.

Since the seed pulses are subject to the same phase fluctuation Δφ as the laser fundamental, and the frequency-doubled pulses feature a CEP fluctuation of 2Δφ, the signal and idler waves in both OPA stages share the CEP of the pump laser [22]. Difference frequency mixing of any pair of waves including the signal, idler, or laser fundamental is, thus, expected to give rise to inherently phase-locked multi-THz pulses. We mix the laser fundamental with the idler output of the OPA because this process only depletes the more scalable laser fundamental whereas the OPA output is amplified. We branch off pulses with an energy of 4.7 μJ from the laser output [branch (ii)]. Since the laser frequency is centered at 292 THz while the OPA idler frequency (wavelength) is tunable between 280 THz (1070 nm) and 250 THz (1200 nm), the difference frequency can be set between 12 and 42 THz. Phase-matched DFG is possible throughout this spectral range using GaSe [3-7,9-11] or LiGaS$_2$ (LGS) [12,13,25] as nonlinear medium. Both crystals are excellent materials for broadband THz generation due to their large nonlinear coefficient, broadband infrared transparency, and strong birefringence that allows for widely tunable phase matching. We choose a 1-mm-thick LGS crystal (cut for DFG with type-II phase matching in the XY-plane) for the generation of pulses centered above 28 THz, while a 200-μm-thick GaSe crystal allows for DFG at lower frequencies, due to its relatively low phonon frequencies.

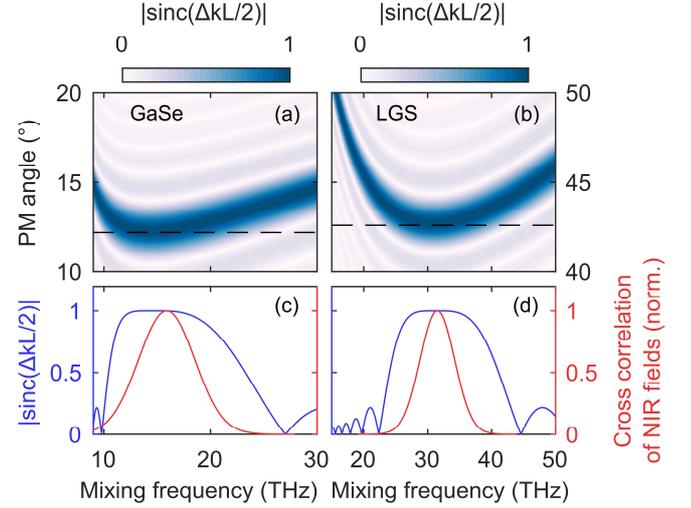

**Fig. 2.** Phase matching (PM) in GaSe and LGS. (a), (b) PM function |sinc(ΔkL/2)| (color coded) as a function of the PM angle and the mixing frequency for type-II DFG in a 200-μm-thick GaSe crystal (a) and a 1-mm-thick LGS crystal (b) for a pump wavelength of 1026 nm. Black dashed lines indicate cross sections shown in (c) and (d). (c),(d) PM curves for perfect group velocity matching (blue curves) and cross correlation between the driving NIR pulses (red curves).

Figures 2(a) and (b) depict the phase-matching function |sinc(ΔkL/2)| as a function of the phase-matching angle and the difference frequency for the two crystals. Here, the phase-matching angle corresponds to the internal propagation angle relative to the Z-axis in GaSe, and relative to the X-axis in LGS. The stationary points located at a frequency of 15 THz in GaSe and 31 THz in LGS represent perfect group-velocity matching between the seed and THz pulses. Under the respective angles [Figs. 2(c) and (d)], the phase-matching bandwidth is determined by the group velocity dispersion [11]. For the relatively low NIR driving frequencies used here, the group velocity dispersion is low, and the maximal phase-matching bandwidth [Figs. 2(c) and (d), blue curves] of our mixing crystals substantially exceeds the maximal multi-THz bandwidth set by the present NIR spectra [Figs. 2(c) and (d), red curves].

The two generating NIR pulses are focused into the mixing crystal to reach an estimated combined peak intensity of 200 GW/cm². The emitted THz radiation is isolated from the NIR pulses by transmission through a germanium filter placed under Brewster's angle. We trace the electric field of these pulses directly by electro-optic sampling with a few-fs gate pulse. For this purpose, part of the laser fundamental [branch (iii), 0.5 µJ] generates a white-light continuum in a 2 mm YAG crystal [25]. A smooth spectrum ranging from 770 to 890 nm is selected from the supercontinuum and compressed in a prism sequence.

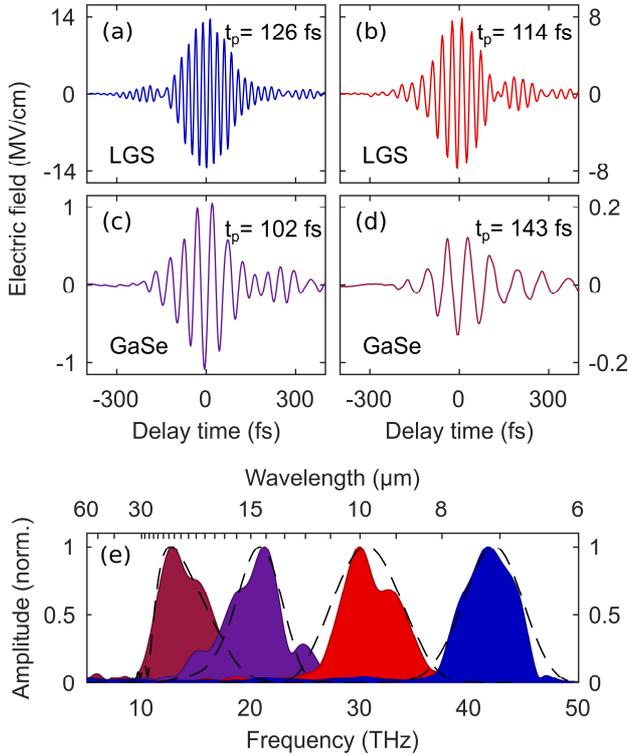

**Fig. 3:** Phase-locked multi-THz waveforms. (a-d) Waveforms generated in a 1-mm LGS crystal (a, b), and a 200-µm GaSe crystal (c, d), retrieved by electro-optic detection. All field transients have been corrected for the electro-optic response function. The pulse duration $t_p$ (intensity FWHM) is denoted next to each transient. (e) Corresponding normalized amplitude spectra obtained by Fourier transforming the corresponding waveforms in (a-d) (filling colors correspond to the respective waveforms). Dashed lines: amplitude spectra calculated by convolving the two NIR pulses and accounting for phase matching.

The resulting 9.3-fs pulse (intensity FWHM retrieved by frequency-resolved optical gating) is used as the electro-optic gating pulse. It is focused collinearly with the multi-THz radiation into a ⟨110⟩-cut ZnTe sensor of a thickness of 6.5 µm contacted to a 300-µm-thick ⟨100⟩-cut ZnTe substrate to sample the waveforms of the multi-THz pulses.

Figure 3 depicts typical field transients generated in various mixing crystals under different phase-matching angles, as detected by ultrabroadband electro-optic sampling. In all cases, the existence of well-defined waveforms clearly proves the CEP-stability of the THz pulse trains emerging from the LGS [Figs. 3(a) and (b)] and GaSe [Figs. 3(c) and (d)] crystals. The corresponding Fourier spectra [Fig. 3(e)] cover the frequency range from 10 to 46 THz, corresponding to a wavelength range between 30 and 6.5 µm. A numerical simulation [Fig. 3(e), broken curves] faithfully reproducing the experimental spectra [Fig. 3(e), solid curves] confirms that the bandwidths are mainly set by the spectra of the NIR driving pulses. The THz pulse duration (intensity FWHM) assumes values between 102 fs [Fig. 3(c)] and 143 fs [Fig. 3(d)]. While the pulse width is typically close to the Fourier transform limit (factor 1.2-1.3), the phonon absorption edge of GaSe leads to a slight dispersion of the lowest frequency components in the transient depicted in Fig. 3(d). Nonetheless, this waveform contains only 1.8 optical cycles within its intensity FWHM.

The average power of the THz pulse train of Fig. 3(a) is measured with a power meter to be as high as 31 mW, corresponding to a pulse energy of 0.16 µJ at 42 THz. Here an excellent quantum efficiency of 24 % with respect to the DFG pump pulses is found. The efficiency with respect to the 1200 nm seed pulses is even 30 %. By measuring both the absolute form of the electromagnetic carrier wave and the Gaussian focal diameter of the THz intensity of 21 µm (FWHM determined with the help of a knife edge and pinholes), we determine a peak electric field of 13 MV/cm – a value comparable to benchmarks that were so far reserved to high-field sources operating at kHz rates.

By carefully adjusting the time delay between the two NIR pulses that drive the DFG process we can readily control the absolute carrier envelope phase $\phi_{THz}$ of the field transients. Figure 4 shows a measured waveform with a frequency centered around 12 THz for three representative values of $\phi_{THz}$. Such precise CEP control is important for non-perturbative nonlinearities, such as high-harmonic generation, occurring on sub-cycle time scales [17].

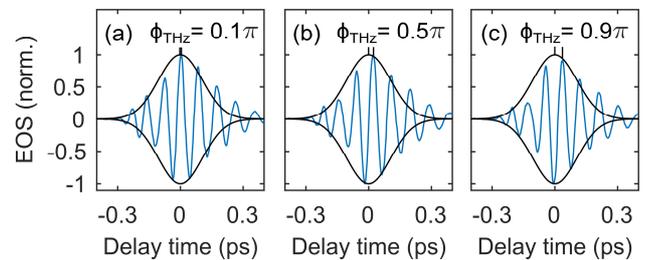

**Fig. 4:** Carrier envelope tunability. Electro-optic trace of multi-THz field transients centered at a frequency of 12 THz for three different values of the carrier envelope phase $\phi_{THz}$ set via the femtosecond delay between the generating NIR pulses.

In order to utilize the possibility of CEP control in practical experiments, however, the pulse train needs to keep a constant phase over multiple hours. The extremely compact one-box architecture of the new source – the two-stage OPA and the MIR generation are hosted in a single box with a footprint of 35 cm × 70 cm – promises excellent long-term stability. A corresponding test is presented in Figure 5, where the same THz waveform is repeatedly measured over the course of 6 hours. In total, 60 traces are taken. We extract $\phi_{THz}$ of each transient by subtracting the delay times of the zero-crossings of the carrier wave from the temporal position of the maximum of the envelope function, calculated via a

Hilbert transformation. The resulting evolution of $\phi_{THz}$ is plotted in Fig. 5(b). From these data, we derive a standard deviation of $\phi_{THz}$ of less than 0.1 π from its mean value. To the best of our knowledge, this result sets a new record in the class of passively CEP-stable OPA-based THz sources [13].

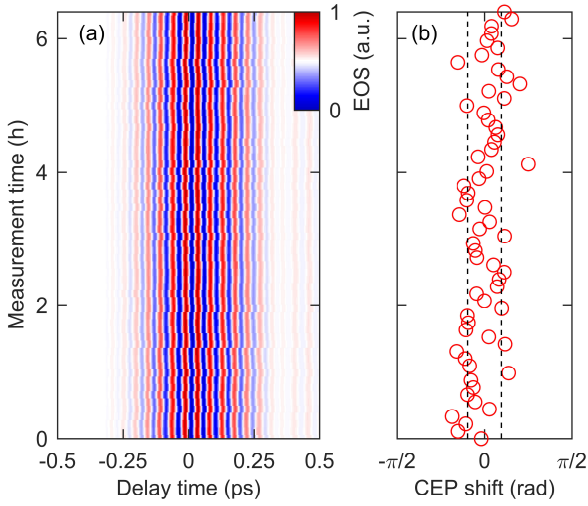

**Fig. 5:** Long-term CEP stability of the multi-THz transients. (a) Electrooptic traces of 60 consecutive measurements of the same field transient generated in a 500-μm-thick GaSe emitter over the course of 6 hours. (b) CEP shift of the transients with respect to their common mean. Black dashed lines indicate the standard deviation of less than 0.1 π.

Our system opens exciting new possibilities for lightwave electronics at high repetition rates. High-harmonic [17] and high-order sideband generation [15,16] driven at repetition rates of 190 kHz and beyond should facilitate multi-octave spanning THz-to-UV frequency combs with average powers exceeding the currently achievable level by orders of magnitude. Intense few-cycle pulses centered at multiple ten THz may also extend lightwave-STM [19,20] to attosecond time scales. Furthermore, the identical CEP of all NIR pulses provides additional flexibility for frequency mixing: For instance, DFG between the signal and idler waves should enable mid-infrared center frequencies of up to 84 THz, whereas multiple DFG branches obtained by separately mixing signal and idler pulses with the laser fundamental may be exploited for two-dimensional multi-THz spectroscopy [26]. By shifting the energy splitting of the pump laser in favor of the DFG pump in the present setup, we should also be able to go from the DFG regime into the OPA regime [22,27,28]. This may lead to a further increase of the THz output while staying safely below the damage threshold of the crystals. Finally, our parametric approach should lend itself naturally to latest power-scaled pump sources. Using a high-power Yb:YAG femtosecond amplifier [29] with a comparable pulse energy of 55 μJ, multi-THz radiation with peak electric fields in excess of 10 MV/cm may be expected at the full repetition rate of up to 20 MHz. 

In conclusion, we employed a second-harmonic pumped OPA concept for the generation of multi-THz pulses tunable from 12 to 42 THz. We reach a very competitive average power with peak fields exceeding 13 MV/cm and pulse durations down to 1.8 cycles at a repetition rate of 190 kHz. Electro-optic sampling with sub-10-fs gate pulses was used for the characterization of the waveforms. Excellent long-term stability was shown resulting from the compact architecture. The high field amplitudes combined with the free tunability and excellent long-term stability of the CEP open the door to the observation of non-perturbative THz nonlinearities at vastly increased repetition rates.

**Funding.** European Union FET-Open Grant (ULTRAQCL 665158) and DFG Cluster of Excellence: Munich-Centre for Advanced Photonics.

**Acknowledgment.** We thank M. Furthmeier for technical assistance.